\documentclass[useAMS,usenatbib,usegraphicx]{mn2e}
\usepackage{psfig}   
\usepackage{graphicx}
\usepackage{subfigure}

\title[The dynamical evolution of very low-mass binaries in open clusters]{\bf The dynamical evolution of very low-mass binaries in open clusters}

\author[R.~J.~Parker and S.~P.~Goodwin]{Richard J.~Parker$^{1,2}$\thanks{E-mail: rparker@phys.ethz.ch} and Simon P.~Goodwin$^2$\vspace*{0.1cm}\\
   $^1$Institute for Astronomy, ETH Z{\"u}rich, Wolfgang--Pauli--Strasse 27, 8093 Z{\"u}rich, Switzerland \\ 
   $^2$Department of Physics and Astronomy, University of Sheffield, 
   Sheffield, S3 7RH, UK}

\begin{document}

\date{}
                             
\pagerange{\pageref{firstpage}--\pageref{lastpage}} \pubyear{2010}

\maketitle

\label{firstpage}

\begin{abstract}
Very low-mass binaries (VLMBs), with system masses $< 0.2$\,M$_\odot$ 
appear to have very different properties to stellar binaries.  This
has led to the suggestion that VLMBs form a distinct and different
population.  As most stars are born in clusters, dynamical evolution
can significantly alter any initial binary population, preferentially
destroying wide binaries.  In this paper we examine the dynamical
evolution of initially different VLMB distributions in clusters to investigate how
different the initial and final distributions can be.

We find that the majority of the observed VLMB systems, which
have separations $<20$~au, cannot be destroyed in even the
densest clusters. Therefore, the distribution of VLMBs with separations 
$<20$~au now must have been the birth population (although we note
that the observations of this population may be very incomplete).  
Most VLMBs with separations $> 100$~au
can be destroyed in high-density clusters, but are mainly unaffected in
low-density clusters.  Therefore, the initial VLMB population must
contain many more binaries with these separations than now, or such
systems must be made by capture during cluster dissolution. 
M-dwarf binaries are 
processed in the same way as VLMBs and so the difference in the
current field populations either points to 
fundamentally different birth populations, or significant observational 
incompleteness in one or both samples.
\end{abstract}   

\section{Introduction}

It has been suggested that brown dwarf-brown dwarf binaries,
or, more generally, very low mass binaries (VLMBs) with system masses
$< 0.2$\,M$_\odot$ form in a different way to stellar
binaries.  The main argument for this scenario is that
the binary fraction and separation distributions of VLMBs are very
different to those of stars.  \citet{Thies07} point out that the
binary fraction of very-low mass systems is only 15 -- 25 per cent,
compared to 42 per cent for higher-mass M-dwarfs.  Also, the separation
distribution  for most stellar binaries (higher-mass M-, K-, and
G-dwarfs) has the same mean (30\,au) and
variance ($\sigma_{\rm log_{10}\,a} = 1.53$, where $a$ is the semi-major axis in au), 
and differs only in the
multiplicity of the primary in the particular mass range
\citep{Duquennoy91,Fischer92}. However, the observed separation
distribution  of VLMBs \citep[see e.g.][]{Burgasser07} show the data
(when fitted with a log$_{10}$-normal) to have a mean of 4.6\,au with
a much smaller variance.  In addition, \citet{Thies07,Thies08} 
argue that the observations of
VLMBs are not consistent with a  continuous IMF over the
hydrogen-burning limit. \citet{Thies07} interpreted this as evidence that VLMBs form
through a different  mechanism to stellar binaries. 

However, it is known that binary populations can undergo 
significant dynamical processing, with many, especially wider 
systems, being destroyed (\citealp{Heggie75, Kroupa95a, Kroupa95b}; \citealp*{Kroupa99};  
\citealp{Kroupa03, Parker09}).  Therefore, the
currently observed binary population, especially in the field, is not
the same as the birth population \citep{Goodwin10}.

In this paper, we
examine to what extent an initial VLMB population can be altered by
dynamical processing and so how different the initial VLMB and stellar
binary populations can be at birth.  We extend the work of \citet{Kroupa03} 
who examine the evolution of a mixed population of
star and very low-mass object (VLMO) binaries and find that,
dynamically at least, VLMBs must form a separate population 
\citep[see also][]{Thies07, Thies08}.  Here we assume that VLMBs are a separate
population and examine what range of initial binary fractions and
separations can reproduce the current observations.  In 
Section~\ref{observe} we review  the available VLMB data, in
Section~\ref{method} we describe the set-up of our  simulations, we
present our results and discussion in
Sections~\ref{results}~and~\ref{discuss};  and we conclude in
Section~\ref{conclusion}. 

\section{Summary of the Observations}
\label{observe}

Data on very low mass binaries (VLMBs) -- binaries with a total system
mass of  $< 0.2$\,M$_\odot$ -- have been collated in the  \emph{Very
  Low Mass Binaries Archive}   \citep[VLMBA,  see][]{Burgasser07}\footnote{See http://www.vlmbinaries.org/ for  an
  up-to-date census of the known very low mass binaries, maintained by
  N.\,\,Siegler, C.\,\,Gelino and A.\,\,Burgasser.}. Given that the
majority of these systems have primary masses  less than the
hydrogen-burning limit ($m_{\rm H} = 0.08$\,M$_\odot$), the VLMBA
data have the potential to provide an excellent constraint on the
hypothesis that  substellar binaries form via a different mechanism or
in a different environment to stellar binaries. 

As of September 2010, the VLMBA lists 99 systems, four of which lack
robust separation  measurements and a further system has a
planetary-mass companion. That leaves 94 systems  to compare with
numerical simulations. 

\subsection{Multiplicity}

We define the multiplicity of VLMBs as
\begin{equation}
f_{\rm VLMB} = \frac{B}{S + B}, 
\end{equation}
where $B$ is the number of binary systems and $S$ is the number of
single systems. We  ignore triple and higher-order systems for the
remainder of this paper. 

The overall multiplicity of VLMBs is open to debate. Based on
potentially undiscovered systems,  \citet{Basri06} suggest a a value
of 0.26 $\pm$ 0.10. This would argue in favour of the VLMBs as a
continuous  population, as the multiplicity of M-dwarfs and G-dwarfs
is 0.42 and 0.58, respectively  \citep{Fischer92,Duquennoy91} 
possibly indicating a smooth decrease in multiplicity
with decreasing primary mass.  However, \citet{Thies07} argue that 
the VLMBA data is consistent with
an overall  multiplicity of 0.15, representing a distinct cut-off from
the stellar binary regime.  

\subsection{Separation distribution}

\begin{figure}
\begin{center}
\rotatebox{270}{\includegraphics[scale=0.4]{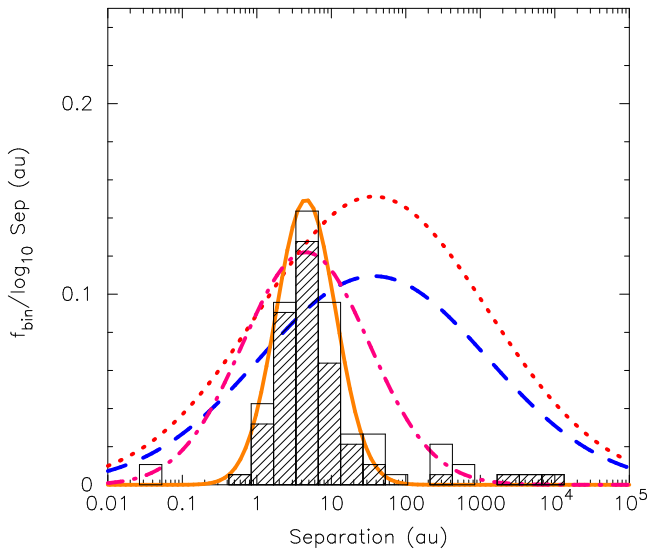}}
\end{center}
\caption[bf]{Data from the Very Low Mass Binary Archive 
\citep[VLMBA --][]{Burgasser07}. The hashed histogram represents binaries 
observed in the Galactic field; whereas the open histogram represents the 
few VLMBs observed in various star clusters. The total area of the histogram 
corresponds to the observed binary fraction of VLMBs, 0.15. The log$_{10}$-normal fits 
to the data by \citet[][the (solid) brown line]{Thies07} and 
\citet[][the (dot-dashed) magenta line]{Basri06} are shown. For comparison, 
the log$_{10}$-normal fits for field M-dwarfs 
\citep[][the (dashed) blue line]{Fischer92} and field G-dwarfs 
\citep[][the (dotted) red line]{Duquennoy91} are also shown.}
\label{VLMBA_sepdata}
\end{figure}

In Fig.~\ref{VLMBA_sepdata} we plot the separation distribution of the
observed  VLMBs, distinguishing between systems observed in the
Galactic field (the  hashed histogram) and systems observed in various
clusters  (the open histogram). The majority of these systems lie in a
small separation range  ($\sim 1-30$\,au with a peak around 5\,au). 
The total area of the histogram corresponds to the observed  
binary fraction of VLM objects, 0.15 \citep{Close03}.

It is important to note that most of the observed VLMBs are in the
field.  This means that their birth cluster has disrupted and that we
are mostly observing an already dynamically processed VLMB
population.  Unfortunately we have no information on the birth
clusters of these systems.

Several authors have attempted to fit the VLMB data with 
log$_{10}$-normal distributions similar to those that are used to fit G-, K- and
M-dwarfs in the field
\citep[][respectively]{Duquennoy91,Mayor92,Fischer92}\footnote{Whilst
  the fit to the G-dwarf data is  log$_{10}$-normal, the K- and
  M-dwarf separation distributions suffer from poorer statistics and
  may  not be a straightforward scaling down of the G-dwarf
  distribution as a function of the multiplicity of the primary
  component (as multiplicity decreases as a function of primary mass in 
  binaries in the field).}.  \citet{Thies07} fit the VLMB separation distribution
with a log$_{10}$-normal distribution with mean 4.6\,au  and variance
$\sigma_{\rm log_{10}\,a} = 0.4$, and an overall substellar
multiplicity of 0.15 (the (solid)  brown line in
Fig.~\ref{VLMBA_sepdata}).

However, there are some outlying systems with very short and long
separations, and \citet{Basri06} argue for  a wider distribution,
based on the hypothesis that there may be unresolved short-period
VLMBs with separations  less than 1\,au \citep{Maxted05} and (at the
time) the tentative discovery of VLMBs with separations in  excess of
100\,au \citep[e.g.][]{Close03,Bouy06}. 

The former is still the subject of debate, with claims that the 
peak in the VLMB separation distribution may be between 1 -- 3\,au 
\citep{Burgasser07}, though \citet{Joergens08} suggests that few VLMBs exist with separations                                                                         $<$\,3\,au. The latter appears to be partly vindicated by recent discoveries of  wider systems in the field
(\citealp[K{\"o}nigstuhl-1 AB, $a = 1800$ au,][]{Caballero07};
\citealp[2M0126AB, $a = 5100$ au,][]{Artigau07};  \citealp[2M1258AB,
  $a = 6700$ au,][]{Radigan09}), although surveys should be sensitive to 
VLMBs with separations between 10 -- 200\,au \citep{Burgasser07}. 

\citet{Basri06} proposed a wider
log$_{10}$-normal fit to the data with mean 4.6\,au and variance
$\sigma_{\rm log_{10}\,a} = 0.85$, and an overall substellar
multiplicity of 0.26 (the (dot-dashed)  magenta line in
Fig.~\ref{VLMBA_sepdata}). For comparison, in Fig.~\ref{VLMBA_sepdata}
we also show the  log$_{10}$-normal fits for field M-dwarfs
\citep[][the (dashed) blue line]{Fischer92} and field G-dwarfs
\citep[][the (dotted) red line]{Duquennoy91}. Details of the
parameters for the  log$_{10}$-normal fits are given in
Table~\ref{lognormfits}.

\begin{table}
\begin{center}
\caption[bf]{Parameters for the log$_{10}$-normal fits to the data in
  Fig.~\ref{VLMBA_sepdata}. The  columns contain, from left to right,
  the reference, mean, log of the mean, variance and the binary
  fraction used to  normalise the distribution. The references are
  \citet[][TK07]{Thies07}, \citet[][BR06]{Basri06},
  \citet[][FM92]{Fischer92} and \citet[][DM91]{Duquennoy91}.}
\begin{tabular}{|c|c|c|c|c|}
\hline
Ref. & Mean & Mean  & Variance & $f_{\rm mult}$ \\
     & (au) & ($\overline{\rm log_{10}\,a}$) & ($\sigma_{\rm log_{10}\,a}$) & \\
\hline
TK07 & 4.6 & 0.66 & 0.40 & 0.15 \\
BR06 & 4.6 & 0.66 & 0.85 & 0.26 \\
FM92 & 30 & 1.57 & 1.53 & 0.42 \\
DM91 & 30 & 1.57 & 1.53 & 0.58 \\
\hline
\end{tabular}
\label{lognormfits}
\end{center}
\end{table}

\begin{figure}
\begin{center}
\rotatebox{270}{\includegraphics[scale=0.4]{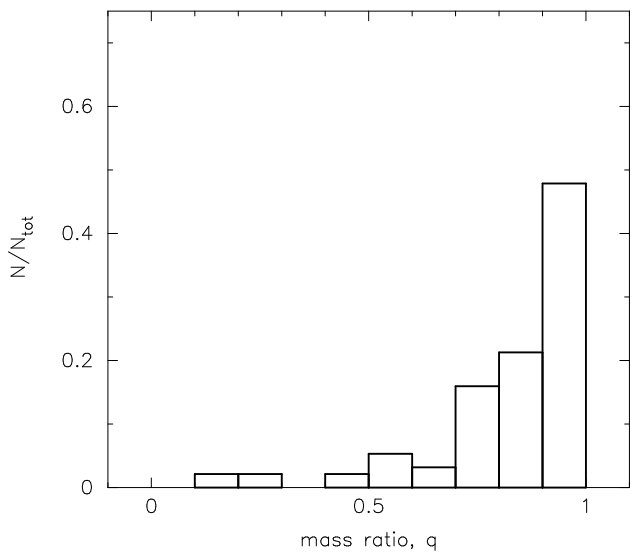}}
\end{center}
\caption[bf]{The mass ratio distribution of systems from the Very Low Mass Binary Archive 
\citep[VLMBA --][]{Burgasser07}. The bins are of width 0.1, and are normalised to the total 
number of systems used (94) to allow direct comparison with the simulations.}
\label{VLMBA_qdata}
\end{figure}

\subsection{Mass ratio distribution}

It is also interesting to trace the possible evolution of the mass
ratio  distribution. For each system, the mass ratio, $q$, is defined
as
\begin{equation}
q = \frac{m_{\rm s}}{m_{\rm p}}, 
\end{equation}
where $m_{\rm p}$ and $m_{\rm s}$ are the masses of the primary and
secondary  components, respectively.

In Fig.~\ref{VLMBA_qdata} we show the observed mass ratio
distribution of the VLMBA data, normalised to the total number of
systems (94).  Almost half the VLMBs in the sample have a mass ratio
approaching unity, and   the majority of the other systems have high
($>$\,0.7) values of $q$.

\subsection{Other properties}

Data on other dynamical properties of the VLMBs are not yet
available. Therefore, we do not  include a study on e.g. the possible
effects of cluster evolution on the VLMB eccentricity  distribution.

\section{Method}
\label{method}

\subsection{Cluster set-up}

We follow a similar method to the one described in \citet{Parker09} to
set up  the clusters and \emph{stellar binary}\footnote{From this
  point in the paper,  we adopt the phrase ``stellar binary'' when
  describing systems with component  masses both exceeding
  0.106\,M$_\odot$; and ``very low-mass binary (VLMB)'' when
  describing systems with both component masses less than
  0.106\,M$_\odot$.  0.105\,M$_\odot$ is the mass of the most massive
  VLMB primary component in the archive.}  systems in our
simulations. The clusters are designed  to mimic a `typical' star
cluster, similar to Orion with $N = 2000$ members and mass  
$\sim 10^3$\,M$_\odot$.

For each set of initial conditions, we create a suite of 10
simulations,  corresponding to 10 clusters, identical apart from the
random number seed used  to initialise the simulations.

We set our clusters up as initially virialised Plummer spheres
\citep{Plummer11} as described by \citet*{Aarseth74}. The prescription
in \citet{Aarseth74} provides the positions and velocities of the
centres of  mass of the systems in the Plummer sphere.

The current half-mass radius of Orion is 0.8\,pc
\citep{McCaughrean94,Hillenbrand98,Kohler06}.  However,
\citet{Parker09} argue that Orion was originally much denser than it
is now  and that the effects of gas expulsion
\citep*{Tutukov78,Hills80,Goodwin97,Kroupa01a,Goodwin06}  and
dynamical interactions (\citealp{Kroupa99}, 
\citealp*{Berk07}; \citealp[][and references
  therein]{Parker09})  have caused it to expand to its current
size. We therefore adopt initial half-mass radii of  0.1\,pc and
0.8\,pc for the clusters in our simulations, thereby covering a wide
range of  cluster densities. 

Observations suggest that the ratio of stars with masses
$<$\,1\,M$_\odot$ to  brown dwarfs is $\sim$\,5:1
\citep[e.g.][]{Andersen08}. In our simulations,  we place one
sub-stellar system (either single or binary) in the cluster
for every five stellar systems.

\subsection{Stellar binary properties}

It is thought that the star formation process should produce binary
stars in  preference to singles \citep[][and references
  therein]{Goodwin05,Goodwin07}.  Therefore, all the clusters in our
simulations are formed with an initial  stellar binary fraction,
$f_{\rm stellar} = 1$ (i.e. all stars form in  binary systems; there
are no singles or triples, etc.), where
\begin{equation}
f_{\rm stellar} = \frac{B}{S + B},
\end{equation}
and $S$ and $B$ are the numbers of single and binary systems,
respectively. 

The mass of the primary star is chosen randomly from a
\citet{Kroupa02} IMF of  the form
\begin{equation} 
 N(M)   \propto  \left\{ \begin{array}{ll} M^{-1.3} \hspace{0.4cm} m_1
   < M/{\rm M_\odot} < m_2   \,, \\ M^{-2.3} \hspace{0.4cm} m_2 <
   M/{\rm M_\odot} < m_3   \,,
\end{array} \right.
\end{equation}   
where $m_1$ = 0.106\,M$_\odot$, $m_2$ = 0.5\,M$_\odot$, and  $m_3$ =
50\,M$_\odot$. Note that the lower mass limit, $m_1$, is  higher than
in our previous papers \citep[e.g.][]{Parker09c,Parker09}. This is to prevent  a
stellar binary or single star from having mass components ($m_p$
\emph{and} $m_s$) that would overlap with the VLMBA data\footnote{We note that in their 
work,  \citet{Thies07} deliberately allowed an overlap in the mass range 0.08 -- 0.1\,M$_\odot$, 
so that objects could be either stars or brown dwarfs.}. In several
test simulations we found that  varying $m_1$ by a few per cent causes
a negligible difference to the results.  

Secondary masses are drawn from a flat mass ratio distribution with
the  constraint that if the companion mass is $<$\,0.106\,M$_\odot$ or
the total  mass of the binary, $m_{\rm tot} < $ 0.2\,M$_\odot$ it is
reselected,  thereby removing the possibility of creating VLMBs in the
stellar domain.  Note that this method does not produce the input IMF, exactly due
  to the way that secondaries are chosen.

In accordance with the observations of \citet{Duquennoy91} and 
\citet{Fischer92}, the periods of stellar binary systems are drawn from a 
log$_{10}$-normal distribution of the form
\begin{equation}
f\left({\rm log_{10}}P\right) \propto {\rm exp}\left \{ \frac{-{({\rm log_{10}}P -
\overline{{\rm log_{10}}P})}^2}{2\sigma^2_{{\rm log_{10}}P}}\right \},
\label{log-norm}
\end{equation}
where $\overline{{\rm log_{10}}P} = 4.8$, $\sigma_{{\rm log_{10}}P} = 2.3$ and
$P$ is in days. The periods are then converted to semi-major axes.

Eccentricities of stellar binaries are drawn from a thermal eccentricity
distribution \citep{Heggie75,Kroupa95a,Kroupa08} of the form
\begin{equation}
f_e(e) = 2e.
\label{ecc_therm}
\end{equation}

Binaries with small periods but large eccentricities would  expect to
undergo the tidal circularisation shown in the sample of G-dwarfs  in
\citet{Duquennoy91}. We account for this by reselecting the
eccentricity  if it exceeds the following period-dependent value
$e_{\rm tid}$:
\begin{equation}
e_{\rm tid} = \frac{1}{2}\left[0.95 + {\rm tanh}\left(0.6\,{\rm
    log_{10}}P - 1.7\right)\right],
\label{ecc_tidal}
\end{equation}
with ${\rm log_{10}}\,P$ in days\footnote{A more elaborate `eigenevolution' algorithm, 
which accounts for pre-Main Sequence tidal circularisation and protostellar disk accretion 
is described in \citet
{Kroupa08}. We elect not to use it here, as the disk accretion mechanism in this algorithm also alters the mass ratio 
distribution.}. This ensures that the
eccentricity--period  distribution matches the observations of
\citet{Duquennoy91}. 

\subsection{Very low-mass binary properties}
\label{sub-stellar_prop}

In three sets of simulations, we use the VLMBA data 
\citep[]{Burgasser07} to randomly choose
binaries and  use their masses and semi-major axes as initial values
for substellar  binaries in the cluster. 

We run simulations with semi-major axes drawn from the \citet{Thies07} fit
to the VLMB  separation distribution, with Gaussian parameters of
$\overline{{\rm log_{10}}\,a} = 0.66$, where $a$ is the semi-major
axis in au  (0.66 corresponds to 4.6\,au), and variance $\sigma_{{\rm
    log_{10}}\,a} = 0.4$; and also the  fit by \citet{Basri06}, which
accounts for the outlying binaries in the  VLMB separation
distribution by adopting the same log$_{\rm 10}$-normal peak but
increasing the  variance ($\overline{{\rm log_{10}}\,a} = 0.66$,
$\sigma_{{\rm log_{10}}\,a} = 0.85$). For completeness, we run
simulations  in which the sub-stellar binaries have the same
separation distribution as  the stellar binaries ($\overline{{\rm
    log_{10}}\,P} = 4.8$,  $\sigma_{{\rm log_{10}}\,P} = 2.3$; $P$ in
days).

In each case, we adopt an initial VLMB fraction; either 0.5 
\citep[c.f.\,\,the stellar binaries in the field][]{Duquennoy91}, 0.25 
\citep[to account for potentially undiscovered VLMBs,][]{Basri06}, or 
0.15 (the \citet{Thies07} fit to the observations). 

In the simulations that choose separations from the various  log$_{\rm
  10}$-normal distributions, the masses of the sub-stellar binary
components are chosen by randomly assigning the primary a mass  in the
range 0.01\,M$_\odot < m_p \leq$ 0.106\,M$_\odot$. We then adopt a
flat mass ratio ($q$) distribution to choose the mass of the secondary
component (0.01\,M$_\odot \leq m_s < m_p$). 

This mass range allows a direct comparison between the VLMBA and the
simulations.  We find that reducing the upper limit of the VLMB
primaries to 0.08\,M$_\odot$  has a negligible effect on the results.

The eccentricities are drawn from a thermal eccentricity distribution
and then  tidally circularised
(Eqns.~\ref{ecc_therm}~and~\ref{ecc_tidal}).

A summary of the different mass, semi-major axes and eccentricity
distributions used to create the substellar binaries is given in
Table~\ref{BDBDtable}.

\subsection{$N$-body integration}

By combining the primary and secondary masses of the stellar and VLMBs
with their semi-major axes and eccentricities, the relative velocity
and radial  components of the stars/very low-mass objects in each system are
determined. These are then  placed at the centre of mass and centre of
velocity for each system in the  Plummer sphere.

Simulations are run using the \texttt{kira} integrator in Starlab
\citep[e.g.][and references therein]{Zwart99,Zwart01} and evolved  for
10\,Myr.  We do not include stellar evolution in the simulations.

\begin{table}
\begin{center}
\caption[bf]{A summary of the different substellar binary configurations 
adopted in the simulations. From left to right, the columns show the 
initial cluster half-mass radius, $r_{\rm 1/2}$; the separation distribution (either 
the log$_{\rm 10}$-normal distributions of \citet[][DM91]{Duquennoy91}, 
\citet[][BR06]{Basri06} and \citet[][TK07]{Thies07} or values taken 
directly from the VLMBA); the initial multiplicity; the masses of the 
two components in a system (either randomly selected with a flat mass-ratio 
distribution, or taken from the VLMBA). See Section~\ref{sub-stellar_prop} for full details.}
\begin{tabular}{|c|c|c|c|}
\hline 
$r_{\rm 1/2}$ & Separation Distribution &  $f_{\rm VLMB}$ & Mass Range \\
\hline
0.1\,pc & VLMBA & 0.25 & VLMBA \\
0.8\,pc & VLMBA & 0.25 & VLMBA \\
0.8\,pc & VLMBA & 0.15 & VLMBA \\
\hline
0.1\,pc & TK07 & 0.15 & Random, flat $q$ \\
\hline
0.1\,pc & BR06 & 0.5 & Random, flat $q$ \\
0.1\,pc & BR06 & 0.25 & Random, flat $q$ \\
\hline
0.1\,pc & DM91 & 0.5 & Random, flat $q$ \\
0.8\,pc & DM91 & 0.5 & Random, flat $q$ \\
\hline
\end{tabular}
\label{BDBDtable}
\end{center}
\end{table}

\section{Results}
\label{results}

\subsection{Finding bound binary systems}
 
We use the nearest-neighbour algorithm described by \citet{Parker09} 
\citep[and independently verified by][]{Kouwenhoven10} to determine whether 
a star/very low-mass object is in a bound binary system.

\subsection{The evolution of the VLMB separation distributions}

In this section we discuss the effects of cluster evolution upon the various 
initial separation distributions used to define the VLMB population in the 
clusters.

\subsubsection{The log$_{10}$-normal fit by Thies \& Kroupa}

\begin{figure}
\begin{center}
\rotatebox{270}{\includegraphics[scale=0.4]{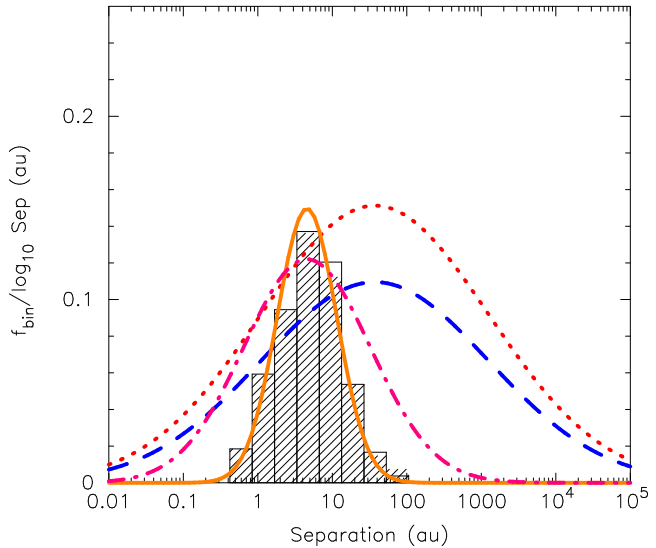}}
\end{center}
\caption[bf]{The evolution of the separation distribution for VLMBs with initial 
separations drawn from the log$_{10}$-normal fit to the observed data by 
\citet[][the (solid) brown line]{Thies07}. The wider fit to the data by
\citet[][the (dot-dashed) magenta line]{Basri06} is shown. The initial VLMB fraction 
is $f_{\rm VLMB} = 0.15$, and the half-mass radius of the cluster is 0.1\,pc. The open histogram 
shows the initial distribution, and the hashed histogram shows the distribution after 1\,Myr.  
For comparison, the log$_{10}$-normal fits for field M-dwarfs 
\citep[][the (dashed) blue line]{Fischer92} and field G-dwarfs 
\citep[][the (dotted) red line]{Duquennoy91} are also shown.}
\label{TK_lognorm_binfrac15}
\end{figure}

In one ensemble of clusters we select separations from the
log$_{10}$-normal fit to  the VLMBA data by \citet{Thies07}. In
Fig.~\ref{TK_lognorm_binfrac15} we show the results of dynamical
evolution in a dense cluster with a half-mass radius of 0.1\,pc. When using the
\citet{Thies07}  log$_{10}$-normal separation distribution as an
initial condition, it is impossible  to dynamically disrupt many of
the systems. \citet{Thies07} used a binary fraction of $f_{\rm VLMB} =
0.15$ to fit the observational data. We show in
Fig.~\ref{TK_lognorm_binfrac15} that the observations can only be
recovered if an  initial binary fraction of $f_{\rm VLMB} = 0.15$ is
used. Any other (higher) fraction hugely  overpopulates the
distribution around its peak, even when adopting a dense cluster
model. 

\begin{figure*}
  \begin{center}
\setlength{\subfigcapskip}{10pt} 
\subfigure[$r_{\rm 1/2}$ = 0.1\,pc]{\label{VLMBA_sep-a}\rotatebox{270}{\includegraphics[scale=0.39]{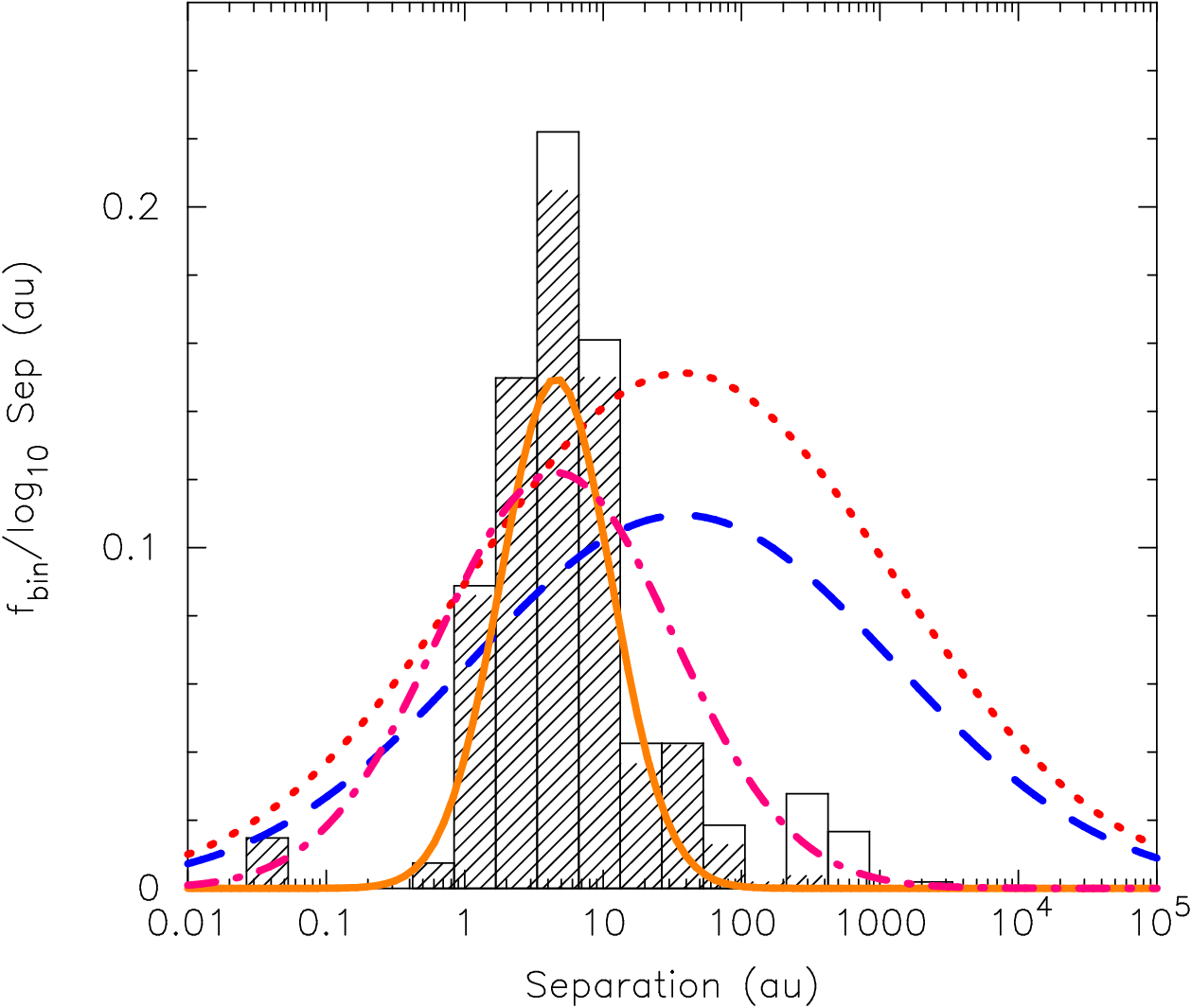}}}
\hspace*{0.6cm} \subfigure[$r_{\rm 1/2}$ = 0.8\,pc]{\label{VLMBA_sep-b}\rotatebox{270}{\includegraphics[scale=0.39]{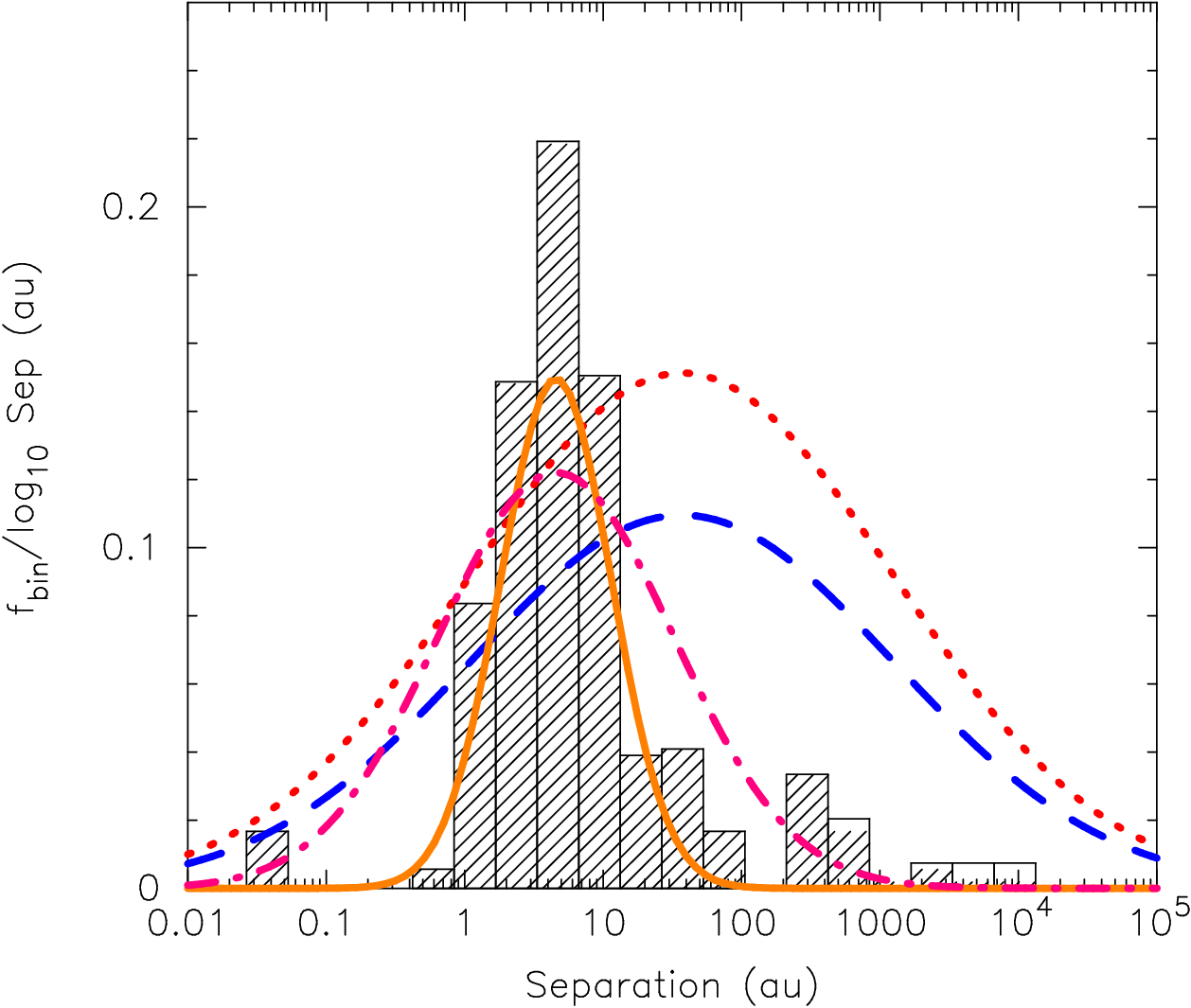}}}
  \end{center}
  \caption[bf]{The evolution of the separation distribution for VLMB with initial 
separations and masses drawn from the VLMBA. The initial VLMB fraction is 
$f_{\rm VLMB} = 0.25$, and the half-mass radius of the cluster is 
(a) 0.1\,pc and (b) 0.8\,pc. The open histogram shows the initial 
distribution, and the hashed histogram shows the distribution after 1\,Myr.
The log$_{10}$-normal fit to the observed data by \citet[][the 
(solid) brown line]{Thies07} and the wider fit to the data by 
\citet[][the (dot-dashed) magenta line]{Basri06} is shown.
For comparison, the log$_{10}$-normal fits for field M-dwarfs 
\citep[][the (dashed) blue line]{Fischer92} and field G-dwarfs 
\citep[][the (dotted) red line]{Duquennoy91} are also shown.}
  \label{VLMBA_sepdists}
\end{figure*}

This initial separation distribution does not account for the wide 
systems, which are observed in both young clusters and in the field. No initial
binary fraction coupled with the \citet{Thies07}  log$_{10}$-normal
fit can reproduce them in any cluster. This is analogous to the result
found by  \citet{Kroupa01b}, who demonstrated that it is impossible to
widen the stellar binary separation distribution through dynamical
evolution in clusters.

\subsubsection{The Very Low Mass Binary Archive data}

In one subset of cluster models, we created the VLMB population by
selecting  systems at random from the online VLMBA. This leads to the
separation distribution shown  in Fig.~\ref{VLMBA_sepdata}.

\begin{figure}
\begin{center}
\rotatebox{270}{\includegraphics[scale=0.4]{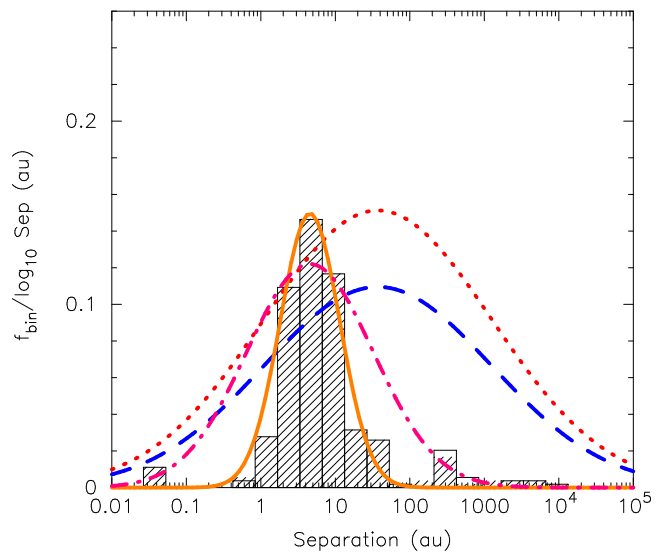}}
\end{center}
\caption[bf]{The evolution of the separation distribution for VLMBs with initial 
separations and masses drawn from the VLMBA. The initial VLMB fraction is 
$f_{\rm VLMB} = 0.15$, and the half-mass radius of the cluster is 0.8\,pc. The open histogram 
shows the initial distribution, and the hashed histogram shows the distribution after 1\,Myr.
The log$_{10}$-normal fit to the observed data by \citet[][the (solid) brown line]{Thies07} 
and the wider fit to the data by \citet[][the (dot-dashed) magenta line]{Basri06} is shown.
For comparison, the log$_{10}$-normal fits for field M-dwarfs 
\citep[][the (dashed) blue line]{Fischer92} and field G-dwarfs 
\citep[][the (dotted) red line]{Duquennoy91} are also shown.}
\label{VLMBA_p8_binfrac15}
\end{figure}

We dynamically evolve three ensembles of clusters containing these
VLMBs. One is a very dense cluster with a half-mass radius  of 0.1\,pc, and
an initial binary fraction $f_{\rm VLMB} = 0.25$. The second is a low-density
cluster with a half-mass radius  of 0.8\,pc (similar to that of Orion
today), and an initial $f_{\rm VLMB} = 0.25$; and the third is a low-density
cluster with a half-mass radius of 0.8\,pc, and an initial $f_{\rm
  VLMB} = 0.15$. The results are shown in
Figs.~\ref{VLMBA_sepdists}~and~\ref{VLMBA_p8_binfrac15}. 

In the dense cluster ($r_{1/2}$ = 0.1\,pc, Fig.~\ref{VLMBA_sep-a}),
the wide VLMBs in the range 100~--~1000\,au  are all destroyed, and
the very wide binaries ($>$\,1000\,au) cannot form at all. This is
because the average distance  between stars in clusters of this
density is $\sim$ 2000\,au, making it highly unlikely that the wide
binaries will be bound  systems, even before dynamical evolution
\citep[see][for a detailed discussion]{Parker09}. There is  some
disruption of the intermediate (4~--~100\,au) VLMBs, but not enough to
drastically  alter the initial distribution. 

When we adopt an initial half-mass radius of 0.8\,pc
(Fig.~\ref{VLMBA_sep-b}), all  the VLMBs placed in the clusters are
found to be bound systems by our algorithm. However,  dynamical
evolution acts to break up the widest ($>$ 5000\,au) VLMBs and systems
with separations less than this are unaffected.  

It is clear that the simulations have too high an initial multiplicity
if we are to match  the observational data. In
Fig.~\ref{VLMBA_p8_binfrac15} we show the $r_{1/2}$ = 0.8\,pc
simulation with a lower initial multiplicity ($f_{\rm VLMB} =
0.15$). This produces the  correct distribution for the intermediate
binaries, but under-produces the number of very wide ($>$\,1000\,au) binaries 
required to be consistent with the observational data.


\subsubsection{The log$_{10}$-normal fit by Basri \& Reiners}

We also conduct simulations in which separations are chosen from the wider
log$_{10}$-normal fit  to the observed data by \citet{Basri06}, in a
dense cluster with a half-mass radius of 0.1\,pc.  The initial
multiplicity is $f_{\rm VLMB}$ = 0.25; we find that any higher
multiplicity  (e.g. $f_{\rm VLMB}$ = 0.5) vastly overproduces the
number of binaries compared to the  observations, even after dynamical
evolution.

\begin{figure}
  \begin{center}
\rotatebox{270}{\includegraphics[scale=0.4]{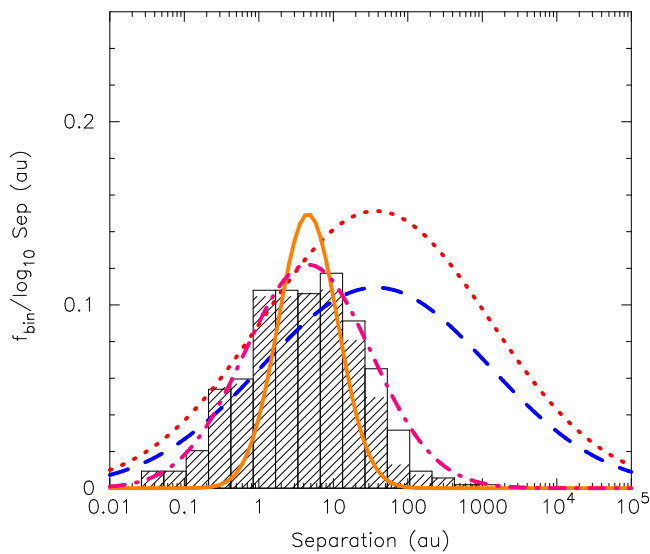}}
  \end{center}
  \caption[bf]{The evolution of the separation distribution for VLMBs 
with initial  separations drawn from the log$_{10}$-normal fit to the
observed data by  \citet[][the
(dot-dashed) magenta line]{Basri06}. The initial multiplicity is
    $f_{\rm VLMB}$ = 0.25 and the initial half mass radius is
    0.1\,pc. The open histogram  shows the initial distribution, and
    the hashed histogram shows the distribution after  1\,Myr. The
    log$_{10}$-normal fit to the observed data by  \citet[][the
      (solid) brown line]{Thies07} is shown. For comparison, the
    log$_{10}$-normal fits for field M-dwarfs \citep[][the (dashed)
      blue line]{Fischer92}  and field G-dwarfs \citep[][the (dotted)
      red line]{Duquennoy91} are also shown.}
  \label{BR_sepdists}
\end{figure}

The resultant separation distribution after 1\,Myr for the cluster
with $f_{\rm VLMB}$ = 0.25  is shown in Fig.~\ref{BR_sepdists}. There
is some dynamical processing of binaries with  separations in excess
of around 10\,au, so that in a dense cluster the \citet{Basri06}
separation distribution is not recovered.  And, like the
\citet{Thies07} separation distribution, this distribution cannot
produce the very wide systems initially (and even if it did they would
be destroyed in a dense cluster).

\begin{figure*}
\begin{center}
\setlength{\subfigcapskip}{10pt}  \subfigure[$r_{\rm 1/2}$ =
  0.1\,pc]{\label{DM_sep-a}\rotatebox{270}{\includegraphics
    [scale=0.39]{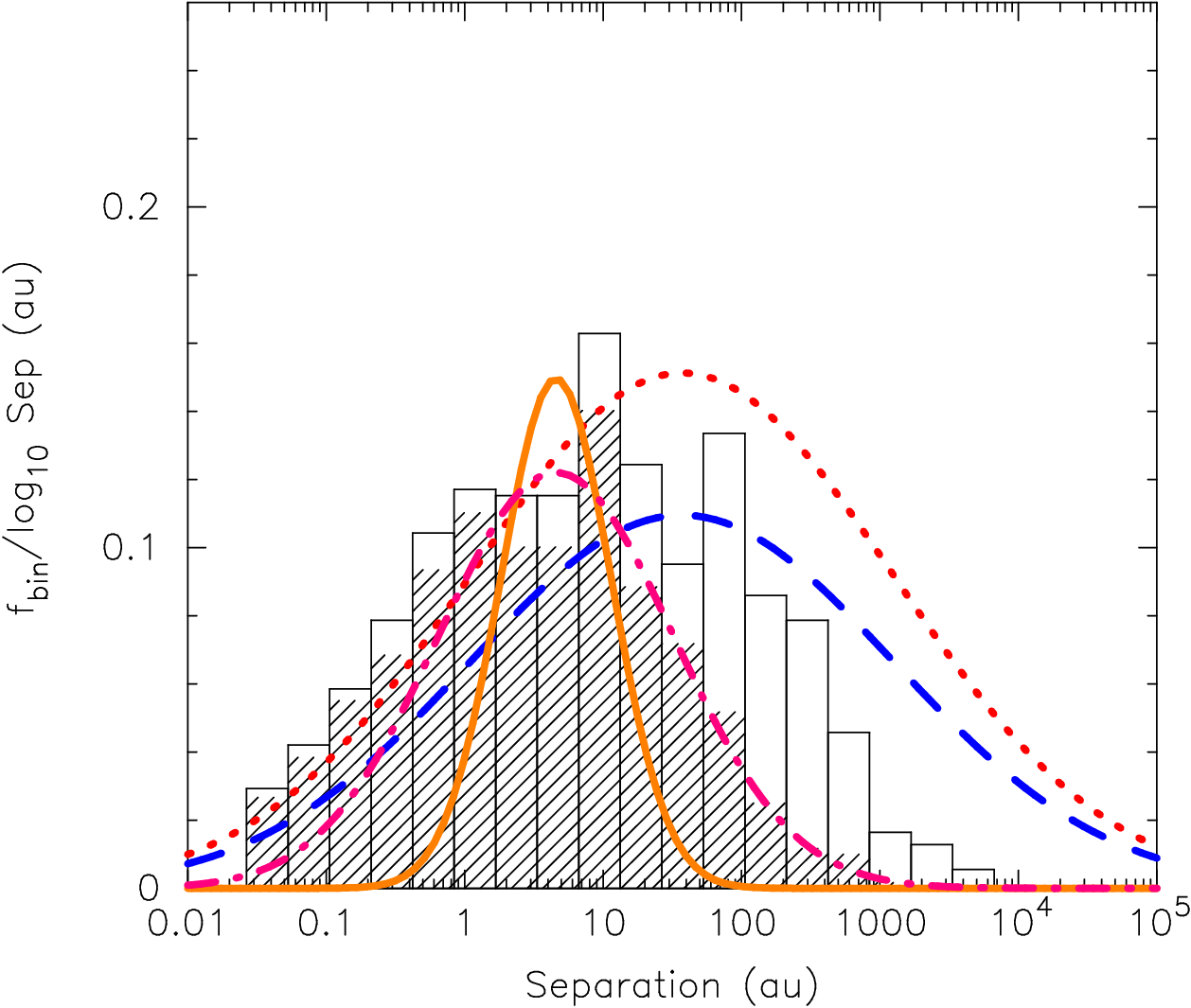}}}
\hspace*{0.6cm} \subfigure[$r_{\rm 1/2}$ =
  0.8\,pc]{\label{DM_sep-b}\rotatebox{270}{\includegraphics[scale=0.39]{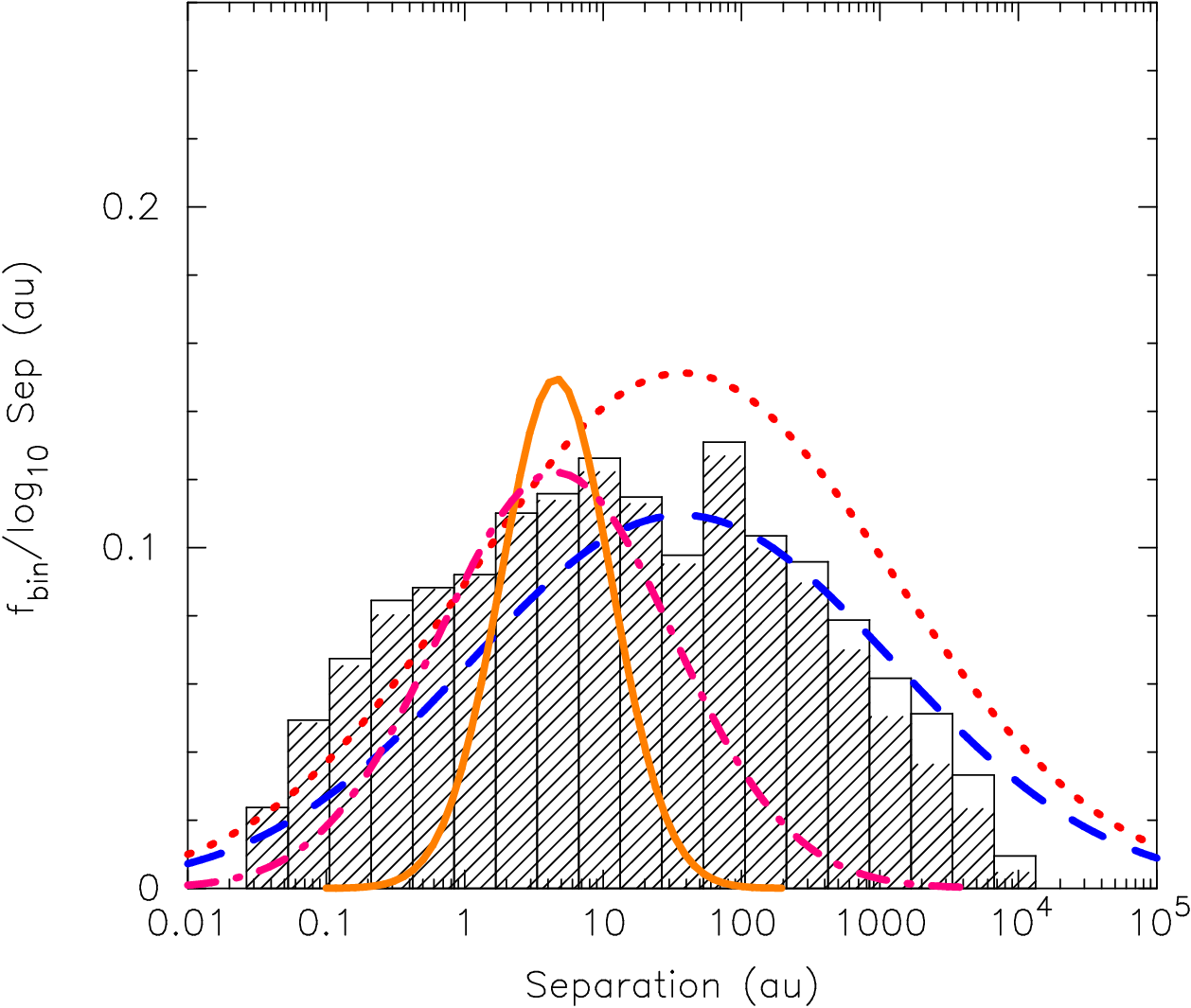}}}
\end{center}
\caption[bf]{The evolution of the separation distribution for
  VLMBs with initial  separations drawn from the same
  log$_{10}$-normal distribution as the stellar binaries. The  initial
  multiplicity of the VLMBs is $f_{\rm VLMBs}$ = 0.5 and we show the
  results  for two different half-mass radii; 0.1\,pc (a) and 0.8\,pc
  (b). The log$_{10}$-normal fits  to the data by \citet[][the (solid)
    brown line]{Thies07} and  \citet[][the (dot-dashed) magenta
    line]{Basri06} are shown. For comparison,  the log$_{10}$-normal
  fits for field M-dwarfs  \citep[][the (dashed) blue line]{Fischer92}
  and field G-dwarfs  \citep[][the (dotted) red line]{Duquennoy91} are
  also shown.}
\label{DM_sepdists}
\end{figure*}

\subsubsection{The log$_{10}$-normal fit for field G-dwarfs}

In the final sub-set of cluster ensembles, the semi-major axes of the
VLMBs are chosen  from the same distribution \citep{Duquennoy91} as
those of the stellar binaries. In Fig.~\ref{DM_sep-a} we  show the
separation distribution of the VLMBs initially (open
histogram),  and after 1\,Myr (hashed histogram) for a dense cluster
($r_{1/2}$ = 0.1\,pc)  with an initial multiplicity of $f_{\rm VLMB}$
= 0.5.

The resultant dynamical processing of the binaries in this dense
environment causes the  separation distribution to evolve into a
\citet{Basri06} log$_{10}$-normal, albeit with  some over-production
of binaries with separations $<$~1\,au. 

In  Fig.~\ref{DM_sep-b} we repeat the simulation, but adopt an initial
half-mass radius  of $r_{1/2}$ = 0.8\,pc for the cluster. With this
initial condition, the cluster is not as  dense, and we overproduce
VLMBs at all separations $>$\,10\,au as dynamical destruction is very
ineffective in such low-density clusters. 

\subsection{Multiplicity fraction}
\label{mult}

\begin{figure}
\begin{center}
\rotatebox{270}{\includegraphics[scale=0.4]{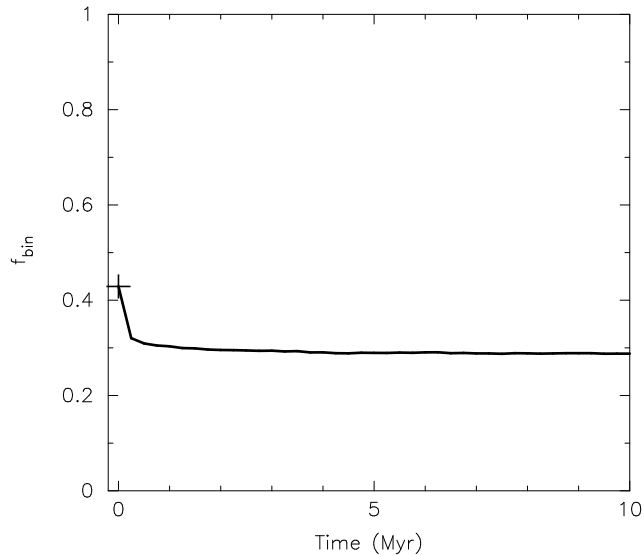}}
\end{center}
\caption[bf]{The evolution of the VLMB fraction over 10\,Myr for
  a cluster  with the VLMB separations drawn from the
  \citet{Duquennoy91} distribution  and an initial binary fraction of
  $f_{\rm VLMB} = 0.5$. The cross shows the initial  binary
  fraction. The initial half-mass radius of the cluster is 0.1\,pc.}
\label{f_multp5}
\end{figure}

\begin{figure}
\begin{center}
\rotatebox{270}{\includegraphics[scale=0.4]{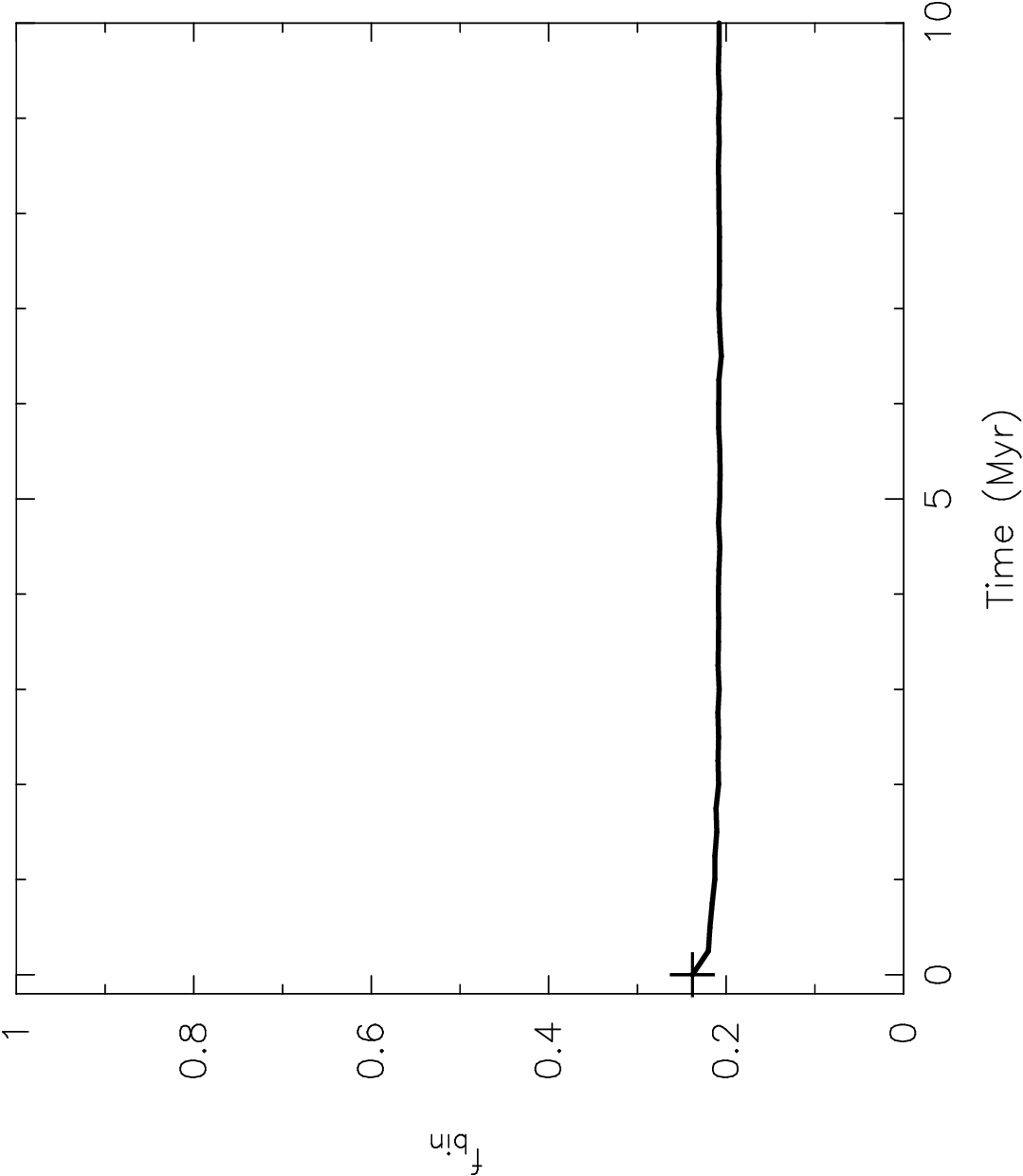}}
\end{center}
\caption[bf]{The evolution of the VLMB fraction over 10\,Myr for
  a cluster  with the VLMB separations drawn from the VLMBA and
  an initial binary fraction  of $f_{\rm VLMB} = 0.25$. The cross
  shows the initial  binary fraction. The initial half-mass radius of
  the cluster is 0.1\,pc.}
\label{f_multp25}
\end{figure}

In each simulation, we determine the multiplicity fraction at each
timestep. For simplicity  (and owing to the very few higher-order
systems that form in the simulations), we ignore  triples, quadruples,
etc.\,\,and simply determine the binary fraction of the VLMBs. We show
the evolution of the VLMB fraction for two of our simulations in
Figs.~\ref{f_multp5}~and~\ref{f_multp25}. 

When the initial binary fraction is 0.5 and separations are drawn from
the  \citet{Duquennoy91} distribution (Fig.~\ref{f_multp5}), there  is
considerable break-up of (wide) systems within the first few crossing
times  within a cluster with an initial half-mass radius of 0.1\,pc
\citep[as detailed for stellar binaries in][]{Parker09}. The initial
binary fraction  is calculated to be 0.42, due to the widest binaries
inputted into the simulations not being  bound within the dense
cluster \citep{Parker09}. The final binary fraction is 0.29,  within
the uncertainty associated with the determination of \citet{Basri06}
of 0.26 $\pm$ 0.10.  If we assume a less dense initial cluster
configuration of 0.8\,pc (to enable the  formation and preservation of
the widest VLMBs observed in the field), then the final  multiplicity
becomes 0.4, which does not agree with the upper limit of the observed
value.  

We also examine the change in binary fraction in a simulation using
the VLMBA  data as initial conditions. During the cluster evolution,
the initially lower  binary fraction of $f_{\rm VLMB} = 0.25$ is also
reduced (Fig.~\ref{f_multp25}),  but only by 0.05. This is due to the
majority of systems being close, and hence  not susceptible to break
up, but also a lower initial binary fraction will be  dominated by the
single very low mass objects -- any break up of the small number  of binaries
will do little to the global binary fraction.

\subsection{The evolution of the VLMB mass ratio distribution}

\begin{figure}
\begin{center}
\rotatebox{270}{\includegraphics[scale=0.4]{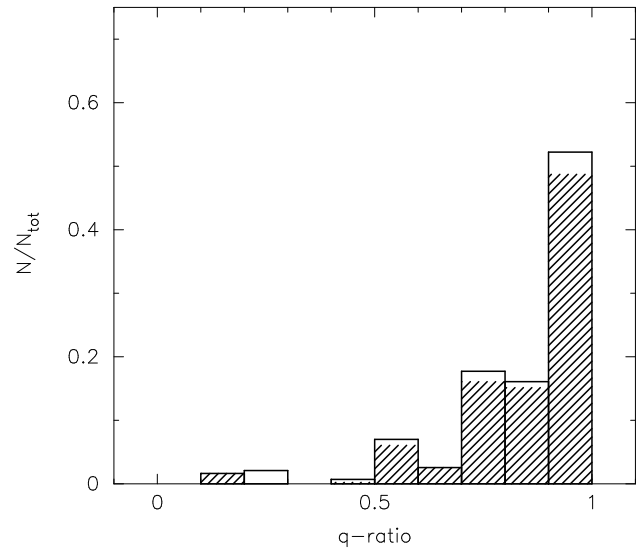}}
\end{center}
\caption[bf]{The evolution of the mass ratio distribution of VLMBs
  with initial  masses and separations drawn from the VLMBA. The
  initial half-mass radius of the  cluster is 0.1\,pc and the data are
  normalised to the number of initial binaries.  The open histogram
  shows the initial distribution, and the hashed histogram shows  the
  final distribution.}
\label{VLMBA_qevol}
\end{figure}

In Fig.~\ref{VLMBA_qevol} we show the effect of cluster evolution on
the mass ratio  distribution of the VLMBs. The results in the plot are
for a cluster with an initial  half mass radius of 0.1\,pc, with
separations and masses chosen from the VLMBA. The  initial
distribution is shown by the open histogram and the final distribution
is  shown by the hashed histogram. The distributions are both
normalised to the total  number of \emph{initial} binaries.

Fig.~\ref{VLMBA_qevol} clearly shows that even in a dense cluster, the
dynamical interactions do not change the mass ratio distribution of VLMBs. 

\section{Discussion}
\label{discuss}

We have examined the dynamical evolution of the binary fractions and
separations of a separate VLMB population in star clusters.  Our aim
is to investigate what the initial VLMB multiplicity and separation
distribution could have have been to reproduce the current
observations. 

Before starting, there are two key points to be considered in this discussion:

Firstly, many binaries are susceptible to destruction in clusters 
\citep[as seen above, also see][]{Heggie75, Kroupa95a, Kroupa95b, Kroupa99, Kroupa03, Parker09}.  
How likely a binary is to be disrupted depends on its separation and
the density of the cluster in which it was born.  In low-density
environments almost all VLMBs can survive, whilst in 
dense clusters VLMBs with separations $>100$~au are susceptible to
disruption (see Fig.~\ref{DM_sepdists}).

Secondly, the field is the sum of star formation from clusters of all
masses and densities, and the field population has already been
dynamically processed.  Therefore, in the field we
are not seeing the birth population of binaries, but a mixed and
evolved population \citep{Goodwin10}.  As the bulk of observed
VLMBs are in the field we are not seeing their birth population.

\subsection{Separation distribution}

Several authors have attempted to parameterise the present-day separation
distribution  of the VLMBs. We set up the VLMBs in our clusters with
four different initial separation  distributions to investigate
whether any of these evolve into the observed distribution,  depending
on the initial density of the cluster (defined by the half-mass radius) 
and multiplicity of the population.

In the simulations in which we use the \citet{Thies07}
log$_{10}$-normal fit to the  data as the initial VLMB separation
distribution, we find that dynamical processing in even the most
dense clusters does not have a significant effect on the initial
binary population. The initial separation distribution does 
not, by construction, contain wide VLMBs, and these are not formed
{\em within} clusters.  However, this distribution could be the
initial VLMB distribution if the very wide binaries are able to form
via capture during the cluster dissolution phase as proposed by 
\citet{Kouwenhoven10} for very wide stellar binaries.  In this
case the initial binarity of the VLMB population must be low at
formation as few binaries are destroyed (and conversely a few very wide
systems may be created).

Using the currently observed VLMBA data as the initial distribution
also does not reproduce the observations.  Again, the close binaries
in the main peak are virtually unaffected by dynamical disruptions.
In this case very wide VLMBs can survive in low-density clusters, 
but they are destroyed in high-density clusters.  Therefore starting
with the observed population cannot reproduce the observed population
unless, once again, some very wide binaries are produced during
cluster dissolution.

Adopting the much wider \citet{Basri06} distribution as an initial
condition has some success in reproducing the observed population in
that it starts with a number of wider VLMBs and so, even after
dynamical processing some wider binaries remain.  However, this
distribution requires us to believe that a large close VLMB population
exists which is currently unobserved \citep{Maxted05,Burgasser07}, though 
see \citet{Joergens08}, and also that a significant
population of VLMBs with separations of tens of~au also remains unobserved, 
which is rather more controversial as direct imaging surveys are sensitive 
to this separation regime in both clusters {\em and} the field \citep{Burgasser07,Joergens08}.

Using the current field initial M-dwarf separation distribution as the initial
VLMB distribution in part fails to match the current VLMB
population.  Whilst high-density clusters can be evolved to
something resembling the \citet{Basri06} distribution, low-density
clusters hardly process their VLMB population at all resulting in far
too many wide- and intermediate-separation VLMBs.

Therefore we can conclude fairly robustly that {\em the currently observed
VLMB fraction and separation distribution are not those of the birth
population}.  This raises the obvious question of what are the binary
fraction and separation distribution of the birth VLMB population?  To
answer this we have to address two inter-related questions.

Firstly, do all clusters produce the same birth populations?  And
secondly, what
is the mixture of cluster densities that contribute to the field?  It
is the combination of cluster densities rather than masses that is
important, as it is density that controls dynamical processing. Any
high-density cluster will process its wide binaries,
whilst a low-density cluster will not, regardless of its mass \citep{Parker09}.

If all clusters do not produce broadly similar birth populations then
it becomes almost impossible to draw many conclusions about birth
populations other than that after dynamical processing they sum to
make the currently observed population (see Goodwin 2010).
However, if we accept as a starting point that star formation is
roughly universal, and that the birth populations of VLMBs in all
clusters are roughly the same then we are able to draw some
conclusions.  

The pertinent points are all illustrated in Fig.~\ref{DM_sepdists}.  In
this figure we see two key points: high-density 
clusters are extremely efficient at destroying VLMBs with separations $>20 - 100$~au,  
whilst low-density clusters leave their birth populations relatively intact; 
and that no cluster can effectively alter the birth populations of VLMBs with 
separations $< 20$~au.  Therefore, the VLMB population with 
separations $< 20$~au must be very close to the birth population, and the
birth population of VLMBs with separations $> 20$~au will have been 
dynamically processed
and must have contained many more systems initially than are observed in the
field.  

This implies that the (putative universal) birth VLMB fraction is
higher than we now observe, and that the birth separation distribution must have a
significant excess of systems with wide separations \citep[see][for a similar 
argument for stellar binaries]{Kroupa95b}.

How much higher the birth binary fraction must have been depends on
the density distribution of star clusters.  It should be noted that
this is not the {\em current} density distribution, but rather the
range of maximum densities clusters have reached during their
evolution.  There are suggestions that many clusters undergo an early
dense phase in which their densities can reach those simulated by the
$0.1$~pc half-mass radius clusters we model here 
\citep[see][]{Kroupa99,Moraux07,Bastian08,Allison09,Parker09,Goodwin10}.  
If many clusters do indeed undergo an
early dense phase, then the initial VLMB population may need to be
similar to the very wide $f_{\rm mult} = 0.5$ population illustrated
in Fig.~\ref{DM_sep-a}.  If, on the other hand, many stars are
created in low-density associations, the initial VLMB population would
only need to contain slightly more wide binaries than are now
observed.

What is clear is that VLMBs and M-dwarfs (taking the \citet{Fischer92} distribution) must have had very different birth populations.
M-dwarfs are subject to exactly the same dynamical processing as VLMBs
(only slightly modified by their larger masses), and to have such 
significantly different separation distributions in the field  means they must
have been born with significantly different distributions.

The key regime in which more observations are required is in
examining if there are more wide and, especially, intermediate
separation VLMBs.  The confirmation of either presence or lack of a 
significant VLMB population with separations of $\sim 100$~au where we
know that there are significant numbers of M-dwarf binaries is crucial
in order to determine how different the initial binary populations are.

\subsection{Mass ratio distribution}

As demonstrated in Fig.~\ref{VLMBA_qevol}, the effect of dynamical
evolution  on the mass ratio distribution of the VLMBs is negligible,
even in the most dense  clusters. What is clear is that there is no
preferential break-up  of systems with a particular $q$ value; the
distribution is uniformly lowered by  dynamical processing, in
agreement with the evolution of the mass ratio distribution  for
low-mass stars ($0.1$\,M$_\odot \leq m_p \leq 1.1$\,M$_\odot$ and
$0.1$\,M$_\odot \leq m_s \leq 1.1$\,M$_\odot$) in clusters found by
\citet{Kroupa03}.  This means that the current mass ratio
distribution represents the birth mass ratio distribution.

\section{Conclusions}
\label{conclusion}

We present the results of $N$-body simulations of the effect of
dynamical evolution on  very low mass binaries (VLMBs) in `typical'
$N = 2000$ Orion-like star clusters with various initial densities. In each
cluster, we place a separate population of VLMBs and follow the
effects of dynamical interactions on these systems. 

Our conclusions can be summarised as follows:

$\bullet$ VLMBs with separations $< 10 - 20$~au cannot be disrupted,
even in the densest clusters.  Therefore the currently observed VLMB
population with separations $< 10 - 20$~au must be the primordial
population.  However, we note that the VLMB population at these
separations is poorly known.

$\bullet$ Many VLMBs with separations $> 100$~au can be destroyed in
very dense clusters (with half-mass radii of $0.1$~pc), but are
relatively unaffected in low-density clusters (with half-mass radii of
$0.8$~pc, so 512 times less dense).

$\bullet$ If all star clusters produce the same {\em birth} VLMB
population, then the birth VLMB binary fraction must have been 
somewhat higher than we observe in the field.  The initial VLMB
distribution must also have significantly more wide VLMBs than are 
currently observed unless capture during cluster dissolution is an
effective wide binary formation mechanism.

$\bullet$ The mass ratio distribution of VLMBs is unchanged by
dynamical processing and so is a probe of the birth mass ratio
distribution.  

$\bullet$ M-dwarf binaries are also processed by clusters in the same
way as VLMBs, and so the very different field populations must reflect
very different birth populations.  

Further work to better constrain (and improve the statistics of) the
separation distribution of both VLMBs and M-dwarf binaries in the field would 
be beneficial
in determining whether the transition in separations between M-dwarf and
VLMBs is as abrupt as Fig.~\ref{VLMBA_sepdata} suggests. If
not, then the trend in decreasing multiplicity  with primary mass, and
fewer wide VLMBs than wide G-dwarf binaries, suggests a possible
continuation through the  hydrogen-burning limit from the stellar to the
substellar regime. 

\section*{Acknowledgements}

RJP acknowledges financial support from STFC. We thank the anonymous referee for 
their comments on the original text, which have greatly improved this paper. This work has 
made use of the 
Iceberg computing facility, part of the White Rose Grid computing facilities 
at the University of Sheffield. This work has also made use of the Very Low Mass 
Binaries Archive maintained by Nick Siegler, Chris Gelino and Adam Burgasser 
at  http://www.vlmbinaries.org.

\bibliographystyle{mn2e}
\bibliography{brown_dwarf_ref}

\label{lastpage}

\end{document}